\documentclass [12pt]{article}
\usepackage[left=2.6cm,top=2.6cm,right=2.6cm,bottom=2cm,nohead]{geometry}
\usepackage{caption}
\usepackage{pdflscape}
\usepackage[round]{natbib}
\usepackage[makeroom]{cancel}
\usepackage{color}
\usepackage{graphicx}
\usepackage{epsfig,amsmath,latexsym,amssymb}
\usepackage{bm}
\usepackage{hyperref}
\usepackage{url}
\usepackage{booktabs}
\usepackage{multirow}
\usepackage{algorithm}
\usepackage{algpseudocode}
\usepackage{rotating}
\usepackage{tikz}

\newtheorem{remark}{Remark}[section]

\begin{document}

\title{
Impact of COVID-19 type events on the economy and climate under the stochastic DICE model}

\author{\small{Pavel V.~Shevchenko$^{1,\ast}$,  Daisuke Murakami$^{2}$, Tomoko Matsui$^{3}$, Tor A. Myrvoll$^{4}$}}

\date{\footnotesize{draft, 1st version, 22 December 2020, revised 9 October 2021}}
\maketitle
\begin{center}
\vspace{-0.65cm}
\footnotesize {\textit{
$^{1}$ Department of Actuarial Studies and Business Analytics, Macquarie University, Australia; e-mail: pavel.shevchenko@mq.edu.au\\
$^{2}$ Institute of Statistical Mathematics, Japan; e-mail:  dmuraka@ism.ac.jp \\
$^{3}$ Institute of Statistical Mathematics, Japan; e-mail: tmatsui@ism.ac.jp \\
$^{4}$ Norwegian University of Science and Technology, Norway; e-mail: tor.andre.myrvoll@ntnu.no\\
$\ast$ Corresponding author}}
\end{center}

\begin{abstract}
\noindent The classical DICE model is a widely accepted integrated assessment model for the joint modeling of economic and climate systems, where all model state variables evolve over time deterministically. We reformulate and solve the DICE model as an optimal control dynamic programming problem with six state variables (related to the carbon concentration, temperature, and economic capital) evolving over time deterministically and affected by two controls (carbon emission mitigation rate and consumption). We then extend the model by adding a discrete stochastic shock variable to model the economy in the stressed and normal regimes as a jump process caused by events such as the COVID-19 pandemic. These shocks reduce the  world gross output leading to a reduction in both the world net output and carbon emission. The extended model is solved under several scenarios as an optimal stochastic control problem, assuming that the shock events occur randomly on average once every 100 years and last for 5 years. The results show that, if the world gross output recovers in full after each event,
the impact of the COVID-19 events on the temperature and carbon concentration will be immaterial even in the case of a conservative 10\% drop in the annual gross output over a 5-year period.
The impact becomes noticeable, although still extremely small (long-term temperature drops by $0.1^\circ \mathrm{C}$), in a presence of persistent shocks of a 5\% output drop propagating to the subsequent time periods through the recursively reduced productivity. If the deterministic DICE model policy is  applied in a presence of stochastic shocks (i.e. when this policy is suboptimal), then the drop in temperature is larger (approximately $0.25^\circ \mathrm{C}$), that is, the lower economic activities owing to shocks imply that more ambitious mitigation targets are now feasible at lower costs.

\vspace{0.5cm}

\end{abstract}

\pagebreak
\section{Introduction}
The impact of the COVID-19 pandemic on the global economy is more severe than the impact from the 2008 global financial crisis (see e.g., \cite{international2020world}), and the projection of the COVID-19 impact on the economy and climate is a major concern.
In this paper, we study the impact of COVID-19 type events on the economy and climate using the dynamic integrated climate-economy (DICE) model, extended to include stochastic shocks to the economy.
The DICE model introduced by Nordhaus\footnote{Nordhaus was the winner of the Nobel Prize in Economic Sciences in 2018 ``for integrating climate change into a long-run macroeconomic analysis".} is an extremely popular integrated assessment model (IAM) for the joint modeling of economic and climate systems. It has been regularly revised over the last three decades with the first version dating back to \cite{nordhaus1992dice}, and the most recent revision being DICE-2016 \cite{nordhaus2017revisiting}\footnote{The most recent version of the DICE model is available at {\url{https://sites.google.com/site/williamdnordhaus/dice-rice}}.}. The DICE model is one of the three main IAMs (the other two are FUND and PAGE) used by the United States government to determine the social cost of carbon; see \cite{scc2016technical}. It balances parsimony with realism and is well documented with all published model equations; in addition, its code is publicly available, which is an exception rather than the rule for IAMs. At the same time, it is important to note that IAMs and the DICE model in particular have significant limitations (in the model structure and model parameters), which have been criticized and debated in the literature (see the discussions in \cite{ackerman2009limitations,pindyck2020use,grubb2021modeling,weitzman2011fat}). Despite the criticism,
the DICE model has become the iconic typical reference point for climate-economy modeling, and is therefore used in our study.

The DICE model is a deterministic approach that combines a Ramsey-Cass-Koopmans neoclassical model of economic growth (also known as the Ramsey growth model) with a simple climate model. It involves six state variables (atmospheric and upper and lower ocean carbon concentrations; atmospheric and lower ocean temperatures; and economic capital) evolving in time deterministically, two control variables (savings and carbon emission reduction rates) to be determined for each time period of the model, and several exogenous processes (e.g., population size and technology level). Uncertainty about the future of the climate and economy is then typically assessed by treating some model parameters as random variables (because we do not know the exact true value of the key parameters) using a Monte Carlo analysis (see \cite{nordhaus2018projections,ackerman2010fat}).

Modeling uncertainty owing to the stochastic nature of the state variables (i.e., owing to the process uncertainty that is present even if we know the model parameters exactly) requires the development and solution of the DICE model as a dynamic model of decision-making under uncertainty, where we calculate the optimal policy response, under the assumption of continuing uncertainty throughout the time frame of the model. This is a much more difficult problem that requires more
computational and mathematical sophistication, whereas the deterministic DICE model can be solved using an Excel spreadsheet or GAMS (a high-level programming language for mathematical modeling {\url{https://www.gams.com/}}).


Few attempts have been made to extend the DICE model to incorporate stochasticity in the underlying state variables and solve it as a recursive dynamic programming problem. For example, \cite{kelly1999bayesian} and \cite{leach2007climate} formulated the DICE model with stochasticity in the temperature-time evolution and solved this as a recursive dynamic programming problem. These studies are seminal contributions to the incorporation of uncertainty in the DICE model (although their numerical solution approach is difficult to extend to higher dimensional space and time-frequency). \cite{cai2012dsice} formulates DICE as a dynamic programming problem with a stochastic shock on the economy and climate. In addition, \cite{traeger20144} developed a reduced DICE model with a smaller number of state variables, whereas \cite{lontzek2015stochastic} studied the impact of climate tipping points. There are other studies that approached optimal strategies addressing climate change through a simple minimization of the damage function to find the optimal timing for an investment, such as in \cite{conrad1997global, luo2013bite}, which is extremely different from the DICE modeling approach and is not pursued in our paper.

In our study, we extend the DICE model by adding a discrete stochastic shock variable, shifting the economy into a stressed regime owing to events such as COVID-19. This is similar to the model formulation in \cite{cai2012dsice} but with different types of jump processes for the shocks. The economy after our shocks is allowed to recover, whereas the jump shock considered in \cite{lontzek2015stochastic, cai2012dsice} is an irreversible climate tipping point event. One of the scenarios we consider allows for stochastic shocks affecting productivity, which leads to a persistent impact on the economy. This is somewhat similar to the tipping point modeling in \cite{lontzek2015stochastic, cai2012dsice}. However, it is important to note that shocks considered in our paper reduce both the world net output and emission through the shock reduction of the gross output while tipping point models assume shock on the net output and no shock on the emission. Thus our shocks lead to a reduction in a policy stringency while tipping point shocks lead to the opposite effect. In addition, our base model is the more recent version of DICE-2016, whereas \cite{lontzek2015stochastic, cai2012dsice} use older DICE versions.

COVID-19 has spread across the globe, with over 75 million confirmed cases and 1.6 million deaths from December 30, 2019, to December 20, 2020, according to the \emph{Weekly Epidemiological Update} from the World Health Organization on December 22, 2020\footnote{ \url{www.who.int/emergencies/diseases/novel-coronavirus-2019/situation-reports}} (with over 4.6 million new cases and 79,000
deaths since the previous weekly updates). Large amounts of emergency loans are needed around the world to develop therapeutic agents and vaccines, as well as to implement various interventions to prevent the spread of infections, such as ``stay-at-home" policies, and provide financial support for them. However, in recent years, the effects of climate change on global warming have become more serious on a global scale and the ``Paris Agreement" to limit the global temperature increase\footnote{The Paris Agreement is an agreement of over 180 countries to reduce greenhouse gas emissions and limit the global temperature increase by $2^ \circ \mathrm{C}$ (above the pre-industrial levels) by 2100, \cite{ParisAgreement2015}.} was adopted in 2015. This has resulted in more public and private funds being provided for green projects involving renewable energy and energy conservation, as the world works to prevent the effects of global climate change. In a pandemic of infectious diseases such as COVID-19, it is important to consider the economic impacts of both such pandemics and global warming at the same time.

One of the unique features of the COVID-19 pandemic is extreme and widespread disruption to the global economy when compared to other global pandemics, such as the 1918 Influenza Pandemic (``Spanish Flu") or the Hong Kong Flu of 1968\footnote{The real GDP was not significantly affected  during previous global pandemics; see for example GDP data for the United States available from https://www.measuringworth.com}.
On the official government website of the Bureau of Economic Analysis of the United States (see \cite{GrossDomesticProduct}), it was reported on September 30, 2020 that the real gross domestic product (GDP) decreased at an annual rate of 31.4\% in the second quarter of 2020. According to a news report on September 2, 2020 in \cite{JapanRecordGDP2020}, Japan’s April-June GDP is expected to be revised after making an annualized 27.8\% drop on a preliminary basis, which is the largest contraction in the post-World War II period. Though, at the time of revision of this paper, the reported drop of the real GDP in the United States in 2020 (compared to 2019) is only 3.4\% (\url{https://www.measuringworth.com} accessed on 9 October 2021).
The amounts of the stimulus packages released by governments in many countries to limit the human and economic impacts of the COVID-19 pandemic have been unprecedented. The International Monetary Fund policy tracker\footnote{\url{www.imf.org/en/Topics/imf-and-covid19/Policy-Responses-to-COVID-19}} presented a summary of the key economic responses around the world (e.g., the Coronavirus Aid, Relief and Economy Security Act introduced in United States in March 2020 has been estimated at 2.3 trillion USD (around 11\% of the nation’s GDP)).

Although the impact of COVID-19 on the economy, human capital, and well-being in the long run is unknown, the historical experience of global pandemics and global recessions can provide valuable insight. \cite{arthi2021disease} provides an excellent review of the long-run effects on health, labor, and human capital from both historical pandemics and historical recessions. It has been argued that, from a historical perspective, the impact of COVID-19 has been similar to that of the ``Spanish Flu" (1918) in terms of direct effects on the health and well-being of individuals, and similar to the Great Depression (1929-1939) in terms of economic disruption.

In this study, we consider the impact of COVID-19 type events on the DICE model outputs. We reformulate the DICE model as an optimal control problem and solve it using dynamic programming involving six state variables evolving over time deterministically and affected by two controls (emission control and savings rates). We then extend the model by adding a discrete stochastic shock variable to the gross world output to shift the economy into a stressed regime owing to events such as COVID-19, assuming that the economy recovers in full after the stressed period. The extended model is solved as an optimal stochastic control problem under different scenarios for the world gross output drop owing to these shock events.
With the reference to the Great Depression (1929-1939) and ``Spanish Flu" (1918), for our scenarios, we assume that shocks occur on average once during a 100-year period and last for 5 years. In addition, the world economic output during a stressed regime decreases by 5\%-10\%. We note that during the Great Depression, the real GDP in the United States dropped for 6 years (1930-1935) comparing to the pre-depression level in 1929 (averaging to 17\% drop per annum over that period)\footnote{See, for example, real GDP data for the United States available from \url{www.measuringworth.com}}, thus our assumption for the shock magnitude is a bit less conservative.
Under all considered conservative scenarios, the impact of COVID-19 type events on the long-term temperature and carbon concentration appear to be quite small.
The results show that if the world’s gross output recovers in full after each event, the impact of COVID-19 on the temperature and carbon concentration will be immaterial even in the case of a conservative 10\% drop in the annual gross output over a 5-year period. The impact becomes noticeable, although remaining extremely small (i.e., a long-term temperature drop by $0.1^\circ \mathrm{C}$), if the shocks are persistent 5\% drops in productivity, leading to a 5\% drop in output propagating to the subsequent time periods. If the deterministic DICE model policy is applied to the stochastic model (i.e., a suboptimal policy is applied in the case of stochastic shocks), then the drop in temperature will be larger (approximately $0.25^\circ \mathrm{C}$), that is, the lower economic activities owing to the occurrence of a shock imply that more ambitious mitigation targets are now feasible at lower costs, which is qualitatively consistent with the results presented in \cite{meles2020covid}.

The remainder of this paper proceeds as follows. The model is defined in Section~\ref{sec:dice}. Section~\ref{sec:numsol} describes the numerical method used to solve the model. The results are presented in Section~\ref{sec:results}, and some concluding remarks are given in Section~\ref{sec:conclusion}.

\section{DICE model}
\label{sec:dice}
The DICE model maximizes the utility of consumption (social welfare) over an infinite time horizon with a tradeoff between consumption, investment, and $\mathrm{CO}_2$ abatement.
Let $t=0,1,\ldots $ be a discrete time measured in steps of $\Delta$ years (e.g., $t=2$ corresponds to $2\Delta$ years).
Using the DICE-2016 model as the foundation\footnote{ \url{www.econ.yale.edu/~nordhaus/homepage/homepage/DICE2016R-091916ap.gms}}, the stochastic DICE model can be formulated as
\begin{equation}\label{optimisation_problem_eq}
V_0(\bm{X}_0)=\sup_{\bm{\mu},\bm{c}}{\mathbb{E}}\left[\sum_{t=0}^{\infty} e^{-\widetilde{\rho}\Delta t} U(c_t,L_t)\right],
\end{equation}
subject to the state vector $\bm X_t=(K_t,\bm{M}_t,\bm{T}_t,I_t)$ evolving over time\footnote{State variables are affected by controls $\bm\mu,\bm c$. It is standard in the mathematical literature to indicate this by corresponding upperscript of the state vector, however, for simplicity of notation we omit this upperscript.} as
\begin{eqnarray}
&&K_{t+1}=K_t(1-\delta_K)^\Delta+\left[\Delta \times (Q_t(K_t,T_t,\mu_t)-c_t)\right]e^{\epsilon^K_{t+1}},\label{state1_eq}\\
&&\bm{M}_{t+1}=\bm{\Phi}^M \bm{M}_{t}+ \Delta\times (\beta E_t(K_t,\mu_t),0,0)^\prime  e^{\epsilon^M_{t+1}},\label{state2_eq}\\
&&\bm{T}_{t+1}=\bm{\Phi}^T \bm{T}_{t}+\Delta\times \left(\begin{array}{c}
                                           \xi_1 F_t(M^{AT}_t) \label{state3_eq}\\
                                           0 \\
                                         \end{array}
                                       \right)+\bm\epsilon^T_{t+1}, \label{state4_eq}\\
&&I_{t+1}=\mathcal{T}^D(I_t,\epsilon_{t+1}^I).
\end{eqnarray}
Here, superscript $\prime$ denotes a transposition,
$\widetilde{\rho}$ is the utility discount rate\footnote{This is the rate at which the social planner discounts the future utility. In the literature, this is also called the ``rate of pure time preference,” ``subjective discount rate,” ``pure rate of time preference,” or ``welfare discount rate.” }, $\delta_K$ is the annual rate of depreciation of capital, and ($\bm{\Phi}^M$,$\bm{\Phi}^T$,$\beta$,$\xi_1$)\ are parameters for the carbon and temperature transition from $t$ to $t+1$. Other variables and functions are as follows.

\begin{itemize}
\item $U(c_t,L_t)$ is the utility function defined as
$$U(c_t,L_t)= \frac{\Delta L_t}{1-\alpha}\left(\left(\frac{c_t}{L_t}\right)^{1-\alpha}-1\right),$$ where  $\alpha\ge 0$ is the risk aversion parameter ($\alpha=1$ corresponds to a logarithmic utility), and $L_t$ is the world population in billions at time $t$.
\item $\bm{c}=(c_0,c_1,\ldots)$ is the consumption $c_t>0$ and $\bm{\mu}=(\mu_0,\mu_1,\ldots)$ is the carbon emission mitigation rate $\mu_t\ge 0$. Their optimal values were determined by solving the optimization problem (\ref{optimisation_problem_eq}).
\item $K_t$ is the world economic capital (in trillions of USD as of 2010).
\item $\bm{M}_t=(M_t^{AT},M_t^{UP},M_t^{LO})^\prime$ is the carbon concentration (in billions of metric tons) in the atmosphere ($M_t^{AT}$), the upper oceans ($M_t^{UP}$), and the lower oceans ($M_t^{LO}$).
\item $\bm{T}_t=(T_t^{AT},T_t^{LO})^\prime$ is the temperature in the atmosphere ($T_t^{AT}$) and the lower oceans ($T_t^{LO}$) measured in degrees Celsius ($^\circ \mathrm{C}$) above the temperature during the year 1900.
\item $(I_t)_{t\ge0}$ is the random shock process used to model stressed regime in the economy during events such as COVID-19. Depending on the model setup, it may have two or more states. For example, one can set $I_t=0$ corresponding to the normal regime and $I_t=1$ corresponding to the stressed regime. Additional states can be introduced to model situations where the stressed regime has to continue for two or more time periods.
\item ($\epsilon^K_t, \epsilon_{t}^M,\bm\epsilon_{t}^T, \epsilon_{t}^I$) are independent and identically distributed random disturbances for $t=1,2,\ldots$. The random disturbance $\epsilon_{t}^I$ corresponds to the transition probability of the shock process $\Pr[I_{t+1}|I_t]$. Other random disturbances correspond to the uncertainties of the world's net output, carbon concentration, and temperature, and can be modeled using e.g. a Gaussian distribution. In this study, all numerical results are presented for the case in which all random disturbances are set to zero, except for $\epsilon_{t}^I$.
\item $Q_t(K_t,T_t,\mu_t)$ is the world’s net output (output net of the damages and abatement) divided between the consumption and investment, $E_t(K_t,\mu_t)$ is the carbon emission (in billions of tons per year), and $F_t(M^{AT}_t)$ is the radiative forcing, which are modelled as
\begin{align}
&Q_t(K_t,T_t,\mu_t)=\Omega_t(\mu_t,\sigma_t,T_t^{AT}) Y(A_t, K_t, L_t),\label{netoutput_eq}\\
&E_t(K_t,\mu_t)=(1-\mu_t)\sigma_t Y(A_t, K_t, L_t)+E_t^{Land},\label{carbon_emission_eq}\\
&F_t(M^{AT}_t)=\eta\log_2(M_t^{AT}/\widetilde{M}^{AT})+F_t^{EX},
\end{align}
where
\begin{align}
&Y(A_t, K_t, L_t)=(1-\chi(I_t)) A_t K_t^\gamma L_t^{1-\gamma}, \label{production_eq}\\
&\Omega_t(\mu_t,\sigma_t,T_t^{AT})=1-\frac{\sigma_t 550(1-0.025)^t \mu_t^{\theta_2}}{1000\theta_2} - \pi_2 [T^{AT}_t]^{2}.  \label{damage_abatement_factor_eq}
\end{align}
Here,  $\chi(I_t)$ is the impact of COVID-19 type shocks on the Cobb–Douglas production function $\widetilde{Y}(A_t, K_t, L_t)=A_t K_t^\gamma L_t^{1-\gamma}$, $A_t$ is the total productivity factor, and $\Omega_t(\mu_t,\sigma_t,T_t^{AT})$ is the damage abatement cost factor. The damage function as a fraction of the gross output is $\pi_2 [T^{AT}_t]^{2}$; see (\ref{damage_abatement_factor_eq}). The model parameter values and deterministic functions $F_t^{EX}$, $E_t^{Land}$, $A_t$, and $\sigma_t$ are specified in Table \ref{Table_dice_param}.\\
~

\emph{Note that,
$Y(A_t, K_t, L_t)$ is the annual gross world output (output before damage and abatement costs) affected by the shocks $\chi(I_t)$; thus, the shocks affect both the net output $Q_t$ in (\ref{netoutput_eq}) and the carbon emissions $E_t$ in (\ref{carbon_emission_eq}).}
\item The carbon price (USD per ton) is calculated as
$$
P_t=\frac{\partial Q_t/\partial \mu_t}{\partial E_t/\partial \mu_t}\times 1000={550(1-0.025)^t\mu_t^{\theta_2-1}}.
$$
\item The typically quoted \emph{savings rate} output from the DICE model is defined as $(1-c_t/Q_t)$.
\end{itemize}

Given that $(\bm X_t)_{t\ge0}$ is a Markov process, the solution to the stochastic DICE model
(\ref{optimisation_problem_eq}) is a standard optimal stochastic control problem for a controlled Markov process (the transition of $\bm{X}_t$ to $\bm{X}_{t+1}$ is affected by $\mu_t, c_t$). For a good textbook treatment of such problems in finance, see \cite{bauerle2011markov}. This type of problem can be solved using the dynamic programming performed recursively backward in time for $t=N-1,\ldots,0$ through the backward induction Bellman equation:
\begin{equation}\label{dp_eq}
V_t(\bm{X}_t)=\sup_{\mu_t,c_t}\left(U(c_t,L_t)+e^{-\widetilde{\rho}\Delta}\mathbb{E}[V_{t+1}(\bm{X}_{t+1})|\bm{X}_t]  \right),\; \mbox{s.t.}\;V_N(\bm{X}_N)=0,
\end{equation}
and the optimal strategy can be found as
\begin{equation}\label{opt_control_eq}
(\mu_t^\ast(\bm X_t),c_t^\ast(\bm X_t))=\arg\sup_{\mu_t,c_t}\left(U(c_t,L_t)+e^{-\widetilde{\rho}\Delta}\mathbb{E}[V_{t+1}(\bm{X}_{t+1})|\bm{X}_t]  \right).
\end{equation}
Note that, the optimal strategy (optimal decision for the values to be set for carbon emission reduction $\mu_t$ and consumption $c_t$)
depends on the information available at time $t$, that is, depends on the state variable
$\bm{X}_t$. In addition, note that to solve the DICE model under the infinite time horizon numerically, one should use a large enough number of time steps $N$ (it should be confirmed by the sensitivity of the numerical solution that $N$ is sufficiently large and thus its impact on the solution for the period of interest is immaterial).

If the random disturbances and impact from the random shock $\chi(I_t)$ are all set to zero, then the above model is reduced to the standard deterministic DICE-2016. The dynamic programming solution (\ref{dp_eq}) is still valid in this case and can be used to solve the model. Note that the standard DICE-2016 solution is a brute force maximization in (\ref{optimisation_problem_eq}) with respect to ($c_0,\ldots,c_{N-1},\mu_0,\ldots,\mu_{N-1}$) and their constraints simultaneously (a total of 200 parameters plus their constraints when $N=100$).

 In one of the scenarios presented in this paper, to introduce a persistent shock to the gross output $Y_t(A_t,K_t,L_t)$, we consider the total productivity $A_t$ affected by the shock variable $I_t$. Then, $A_t$ becomes an additional state variable, and the new state vector is $\bm{X}_t=(A_t, K_t,\bm{M}_t,\bm{T}_t,I_t)$ with the additional state transition equation,
\begin{equation}\label{Aprod_shock_eq}
A_{t+1}=A_{t}(1+g_A(t))\times (1-\varphi(I_t)),
\end{equation}
where $\varphi(I_t)$ is the impact of shock $I_t$ on productivity, which leads to a persistent shock on the annual economic net output and emission through the above recursive formula. 

\begin{remark}
Utility discounting can be interpreted as the relative weighting given to the well-being of various generations. The choice of an appropriate utility discount rate $\widetilde{\rho}$ is a controversial subject in the literature on global warming models. Some economists have argued that a small or zero utility discount rate $\tilde\rho$ should be used to weight different generations. This topic is
discussed in detail in a Stern Review \cite[Chapter 9]{stern2007economics}. In DICE-2016, the utility discount rate $\tilde\rho$ is set to 1.5\% per year, and the risk-aversion parameter $\alpha$ is 1.45. These parameters are set to generate consumption rates and real returns on capital, consistent with observations (see the discussions in \cite{nordhaus2018projections}).
This approach to setting the discount rate in the DICE model is called a ``descriptive
approach." Under this approach, the real return on capital $r$ is not an exogenous but endogenous variable determined through the Ramsey equation $r=\tilde\rho+\alpha g^\ast$, where $g^\ast$ is the rate of growth of consumption; see \cite[chapter 3]{nordhaus2008question}.
Thus, we assume that the economy shock events do not affect the utility discounting rate, although the real return on capital $r$ implied by the affected consumption and risk aversion can change.
\end{remark}


\begin{table}[!h]
\captionsetup{width=0.95\textwidth}\caption{\footnotesize{The DICE 2016 model parameters, available from \url{www.econ.yale.edu/~nordhaus/homepage/homepage/DICE2016R-091916ap.gms}.}}\label{Table_dice_param}
\begin{center}
\begin{tabular}{l}
  \toprule
  $t=0,\ldots,N, N=99$ with time step $\Delta=5\mbox{ years}$, $t=0$ corresponds to the year 2015\\ \\
 $L_t = L_{t-1}\left(\frac{11.500}{L_{t-1}}\right)^{0.134},\;L_0=7.403\;(\mbox{in billions, } 10^9)$\\
 \\
$A_t=A_{t-1}(1+g_A(t-1)),\;\; g_A(t-1)=\frac{0.076 \exp(-0.005 t\Delta)}{ 1-0.076 \exp(-0.005 t\Delta)},\;\; A(0) = 5.115 $\\ \\
$\sigma_t = \sigma_{t-1} e^{g_{t-1}\Delta},\;\; g_t = g_{t-1}(1-0.001)^\Delta,\;\;    \sigma_0 = \frac{35.85}{105.5(1-0.03)},\;\; g_0 = -0.0152$\\ \\
$E_t^{Land}=2.6(1-0.115)^t$,\;\; $F_t^{EX}= \left(0.5+\frac{t}{34} \right) 1_{t< 17} + 1_{t \ge 17}$\\ \\
  $K_0= 223$,\; $M_0^{AT}=851,\;M_0^{UP}=460,\;M_0^{LO}=1740$,\;
  $T_0^{AT}=0.85,\;T_0^{LO}=0.0068$\\ \\
  $\alpha=1.45,\; \gamma=0.3,\;\rho=0.015, \;\widetilde{\rho}=\ln(1+\rho),\; \delta_K = 0.1$\\ \\
  $\bm{\Phi}^M=\left(
                                                               \begin{array}{ccc}
                                                                 \phi_{11} & \phi_{12} & 0 \\
                                                                 \phi_{21} & \phi_{22} & \phi_{23} \\
                                                                 0  &  \phi_{32} &   \phi_{33}\\
                                                               \end{array}
                                                             \right), \;\;
\bm{\Phi}^T=\left(
                                                               \begin{array}{cc}
                                                                 1-\xi_1\xi_2-\xi_1\xi_3 & \xi_1\xi_3 \\
                                                                 \xi_4 & 1-\xi_4 \\
                                                               \end{array}
                                                             \right)$\\ \\
  $\beta=1/3.666,\; \phi_{21}=0.12,\; \phi_{32}  = 0.007,\; \phi_{11}=1 - \phi_{21},\; \phi_{12}=\phi_{21}588/360$\\
  $\phi_{22}= 1 - \phi_{12} - \phi_{32}, \;\phi_{23} = \phi_{32}360/1720,\;  \phi_{33} = 1 - \phi_{23}$\\
$\xi_4= 0.025,\; \xi_1 = 0.1005,\; \xi_3 = 0.088,\; \xi_2=3.6813/ 3.1$\\ \\
$\eta=3.6813, \; \widetilde{M}^{AT}=588,\; \pi_2=0.00236, \; \theta_2=2.6$ \\
  \bottomrule
\end{tabular}
\end{center}
\end{table}

\section{Numerical solution}
\label{sec:numsol}
The stochastic DICE model can be solved using the Bellman equation (\ref{dp_eq}) and the optimal decisions $\mu_t^\ast(\bm X_t),c_t^\ast(\bm X_t)$ can be found using (\ref{opt_control_eq}) applied backward in time through numerical deterministic dynamic programming (and then if needed we can simulate forward in time random trajectories of $\bm X_t$ based on the calculated optimal decisions to assess the uncertainty). The logical steps of this numerical procedure are presented in Algorithm \ref{dp_algo}. This type of algorithm is often referred to in the literature as a \emph{value function iteration}. Hereafter, $\mathcal{T}(\cdot)$ denotes the transition function for the evolution of state variables
$$\bm{X}_{t+1}=\mathcal{T}(\bm{X}_t,\mu_t,c_t,\bm\epsilon_{t+1})$$
implied by the state processes (\ref{state1_eq}-\ref{state4_eq}), where $\bm\epsilon_{t+1}$ is the vector of random disturbances of the state variables.
Algorithm \ref{dp_algo} is the standard approach for  solving dynamic programming problems numerically. Its performance depends on problem-specific details, such as the type of interpolation across grid points and the type of method used to calculate the required expectations. For example, \cite{cai2012dsice} utilizes Chebyshev nodes for grid points and the Chebyshev polynomial approximation for interpolation. Cubic spline interpolation is also a possible choice. In our numerical experiments, we observed that even the simplest linear interpolation works extremely well for DICE model (it is not the most efficient but is the simplest and quickest way to implement the algorithm).

\begin{algorithm}[h!]
\caption{{Dynamic programming with deterministic grid}}\label{dp_algo}
\begin{algorithmic}[1]
\State Discretize the state variable space to obtain nodes $x_j$, $j=1,\ldots,J$. This discretization can be different for different time slices, $t$. The state variable vector may include discrete and continuous variables (in this case, only continuous variables should be discretized).
\State Initialize $\widehat{V}_N(x_j)=0$ 
for $j=1,\ldots,J$.
\For{${t} = N-1 \,\, \mathbf{to} \,\, 0$}
\State {Interpolate across $\widehat{V}_{t+1}(x_j)$, $j=1,\ldots,J$ to obtain the approximation $\widehat{V}_{t+1}(x)$ for any $x$. This step is not necessary when $t=T-1$ because the maturity condition can be found for any $x$ without interpolation.}
\For{$j=1,\ldots,J$}
\State $\widehat{V}_t(x_j)=\sup_{\mu_t,c_t}\left(U(c_t,L_t)+e^{-\widetilde{\rho}\Delta}\mathbb{E}[\widehat{V}_{t+1}(\widetilde{x})|\bm X_t=x_j]  \right),$
\State where $\widetilde{x}=\mathcal{T}(x_j,\mu_t,c_t,\epsilon_{t+1})$.
\EndFor
\EndFor
\State Simulation of $M$ optimal trajectories forward in time starting from $\bm{X}_0$.
\For{${m} = 1 \,\, \mathbf{to} \,\, M$}
\For{${t} = 0 \,\, \mathbf{to} \,\, N-1$}
\State $(\mu_t^\ast,c_t^\ast)=\arg\sup_{\mu_t,c_t}\left(U(c_t,L_t)+e^{-\widetilde{\rho}\Delta}\mathbb{E}[\widehat{V}_{t+1}(\widetilde{\bm{X}})|\bm X^{(m)}_t]  \right),$
\State where $\widetilde{\bm{X}}=\mathcal{T}(\bm{X}^{(m)}_t,\mu_t,c_t,\bm\epsilon_{t+1})$ and
\State $\widehat{V}_{t+1}(\widetilde{\bm{X}})$ is found by interpolation across $\widehat{V}_{t+1}(x_j)$, $j=1,\ldots,J$.
\State Simulate $\bm\epsilon^{(m)}_{t+1}$ from the model specified distributions.
\State Calculate $\bm{X}^{(m)}_{t+1}=\mathcal{T}(\bm{X}^{(m)}_t,\mu^\ast_t,c^\ast_t,\bm\epsilon^{(m)}_{t+1})$.
\EndFor
\EndFor
\end{algorithmic}
\end{algorithm}

The calculation of $\mathbb{E}[\widehat{V}_{t+1}(\widetilde{x})|\bm{X}_t] $ in Algorithm \ref{dp_algo} can be accomplished through a simulation or quadrature integration methods with respect to the continuous state random variables and simple summation with respect to the discrete state random variables. In our study, we consider only one discrete random variable representing the shock of COVID-19 type events on the world gross output. Thus, the expectation is simply the sum over the states of the shock variable $I_t$. In addition, note that the interpolation on line 4 in Algorithm \ref{dp_algo} is required across grid points of the continuous state variables only.
In the case of many stochastic state variables (i.e., if we want to account for stochasticity in all state variables), we can use the least squares Monte Carlo with control randomization proposed in \cite{Kharroubi2014} with some special adjustments to handle the expected utility problems introduced in \cite{Andreasson2017LSMC}. This goes beyond the purpose of this study and is the subject of our ongoing research project.

For numerical calculations, we implemented Algorithm \ref{dp_algo} in the statistical computing programming language {\tt R}\footnote{https://www.r-project.org/} and then in Fortan because of the long computational times for some scenarios\footnote{Depending on the scenario, the computational time was between 2 and 20 h}. We used the following settings:
\begin{itemize}
\item Each of the six continuous state variables is discretized using equally spaced points (9 points for $K_t$ and 5 points for other variables) and we use a two-state discrete shock variable $I_t$, i.e., in total, there are $2\times 9\times 5^5$ points in the deterministic grid in Algorithm \ref{dp_algo}. We verified that the increase in the number of discretization points does not have a material impact on the results. In the case of productivity $A_t$ affected by shocks (see Eq. (\ref{Aprod_shock_eq}), we discretize $A_t$ using 9 points.
\item To approximate the infinite time horizon, we use $N=80$ (i.e., 400-year time horizon) and then report the results for $t=0,1,\ldots,40$ (i.e., up to 200 years). We verified that increasing the time horizon did not materially change the results.
\item  The range for $K_t$ state variable is selected to be time-varying because this variable changes from $K_0=223$ to approximately $8,000$ at $t=40$. We denote the solution of the standard DICE-2016 model for capital $K_t$ as $\widetilde{K}_t$. The range is then set to $[0.6\widetilde{K}_t,\;1.4 \widetilde{K}_t]$ for $t=0,\ldots,N$.
The ranges for temperatures $T_t^{b}$ ($b=AT,LO$) are set to $[0,\; 1.4 \widetilde{T}_{\max}^{b}]$, and the ranges for carbon concentrations $M_t^{a}$ ($a=AT,UP,LO$) are set to $[0.6\widetilde{M}_{\min}^{a},\;1.4 \widetilde{M}_{\max}^{a}]$. Here, $\widetilde{T}_{\max}^{b}$,  $\widetilde{M}_{\min}^{a}$, and $\widetilde{M}_{\max}^{a}$ are the maximum
temperature, minimum concentration and maximum concentration of the standard DICE-2016 solution, respectively. It was verified that increasing the bounds did not cause any material difference. In the case of stochastically affected productivity $A_t$, the range is set as $[0.6\widetilde{A}_t,\;\widetilde{A}_t]$, where $\widetilde{A}_t$ is a deterministic function of productivity used in the standard DICE-2016 (see Table \ref{Table_dice_param}).
\item The optimal values of the control variables $(\mu_t,c_t)$ are not calculated at $t=0$ but set to the values produced by the standard DICE-2016 because $t=0$ corresponds to the year 2015, which is already in the past. For other time periods, optimal controls are calculated in Algorithm \ref{dp_algo} on lines 6 and 13 using numerical maximization with the same bounds on $\mu_t$ as in the standard DICE-2016.
\item We verified that when stochasticity is set to zero, then our numerical dynamic programming solution leads to virtually the same results as from the original deterministic DICE model.
\item We also set $I_0=0$ and $I_1=1$ for all trajectories in Algorithm \ref{dp_algo} to reflect the fact that there is a shock at the beginning of 2020 and no shock in 2015.
\end{itemize}

To allow for a random change from the normal economy regime to the stressed regime for each time period owing to COVID-19 like events, one could consider the shock variable $I_t$ with two states such that $I_t=0$ corresponds to the normal regime and $I_t=1$ corresponds to the stressed regime. To enforce the change from the stressed to the normal regime, the matrix of transition probabilities $\Pr[I_{t+1}|I_t]$ can be defined as
\begin{equation}\label{trmatrix_eq}
\left( \begin{array}{cc}q&1-q\\1&0\\ \end{array}\right).
\end{equation}
Here, $q$ is the probability of moving from $I_t=0$ to $I_{t+1}=0$, and $(1-q)$ is the probability of moving from $I_t=0$ to $I_{t+1}=1$. If the annual probability of a COVID-19 type event is $p$, then the transition probability $q$ over $\Delta$ years can be approximated as $q = (1-p)^\Delta$.

Note that modeling of climate tipping point shocks such as in \cite{cai2012dsice} can be achieved using the above setup with the important difference; the probability of shock $(1-q)$ should become a function of temperature $T_t^{AT}$ and the second row of the transition probability matrix (\ref{trmatrix_eq}) should be changed from $(1 \quad 0)$ to $(0\quad 1)$. That is, the shock of the tipping point event is irreversible, whereas in the case of COVID-19 type shocks, the stressed regime is forced to be followed by the normal regime. Scenario B calculated and discussed in the next section assumes persistent shocks somewhat similar to the tipping point modeling but note that tipping point modeling assumes shocks to the net output only and no shocks to the emission.


\section{Results}
\label{sec:results}
To study the impact of the COVID-19 type events on the world economy and climate under the DICE model, we calculated the following four scenarios. We set the recovery duration, the decrease in gross world output, and frequency of such events with reference to the Great Depression (1929-1939) for the economic impact and to the ``Spanish Flu" (1918) for the frequency of the events.\\

\noindent \textbf{Scenario A1)} Random shocks reduce the gross world output by 5\% and it takes 5 years to recover in full. That is, in equation (\ref{production_eq}) for gross output, we set $\chi(I_t)=0.05$ if $I_t>0$, and is zero otherwise. These events occur on average once in a 100-year period (i.e., the annual event probability is $p=0.01$). \\

\noindent \textbf{Scenario A2)} Random shocks reduce the gross world output by 10\% and it takes 5 years to recover in full. That is, in equation (\ref{production_eq}) for gross output, we set $\chi(I_t)=0.1$ if $I_t>0$, and is zero otherwise. These events occur on average once in a 100-year period (i.e., the annual event probability is $p=0.01$). \\

\noindent \textbf{Scenario B)} Random shocks reduce productivity $A_t$ by 5\%, i.e., in equation (\ref{Aprod_shock_eq}) we set $\varphi(I_t)=0.05$ if $I_t>0$, and is zero otherwise. We also set $\chi(I_t)$ in equation   (\ref{production_eq}) to be the same as $\varphi(I_t)$  so that gross output persistent drop starts at time $t$. This leads to the persistent drop in the net output and emission.
\\

\noindent \textbf{Scenario C)} The same shock parameters as in Scenario B are used, i.e., $\varphi(I_t)=0.05$ and $\chi(I_t)=0.05$ for $I_t>0$; however, we assume that the control decisions $\mu_t$ and $c_t$ undertaken are the same as in the deterministic DICE model. That is, when simulating trajectories, in Algorithm \ref{dp_algo} on line 13, we do not calculate the optimal stochastic control for the stochastic DICE model but use controls found by deterministic DICE. This also means that we undertake suboptimal decisions.\\

\begin{figure}[p]
\hspace{-1.cm}\includegraphics[scale=0.6]{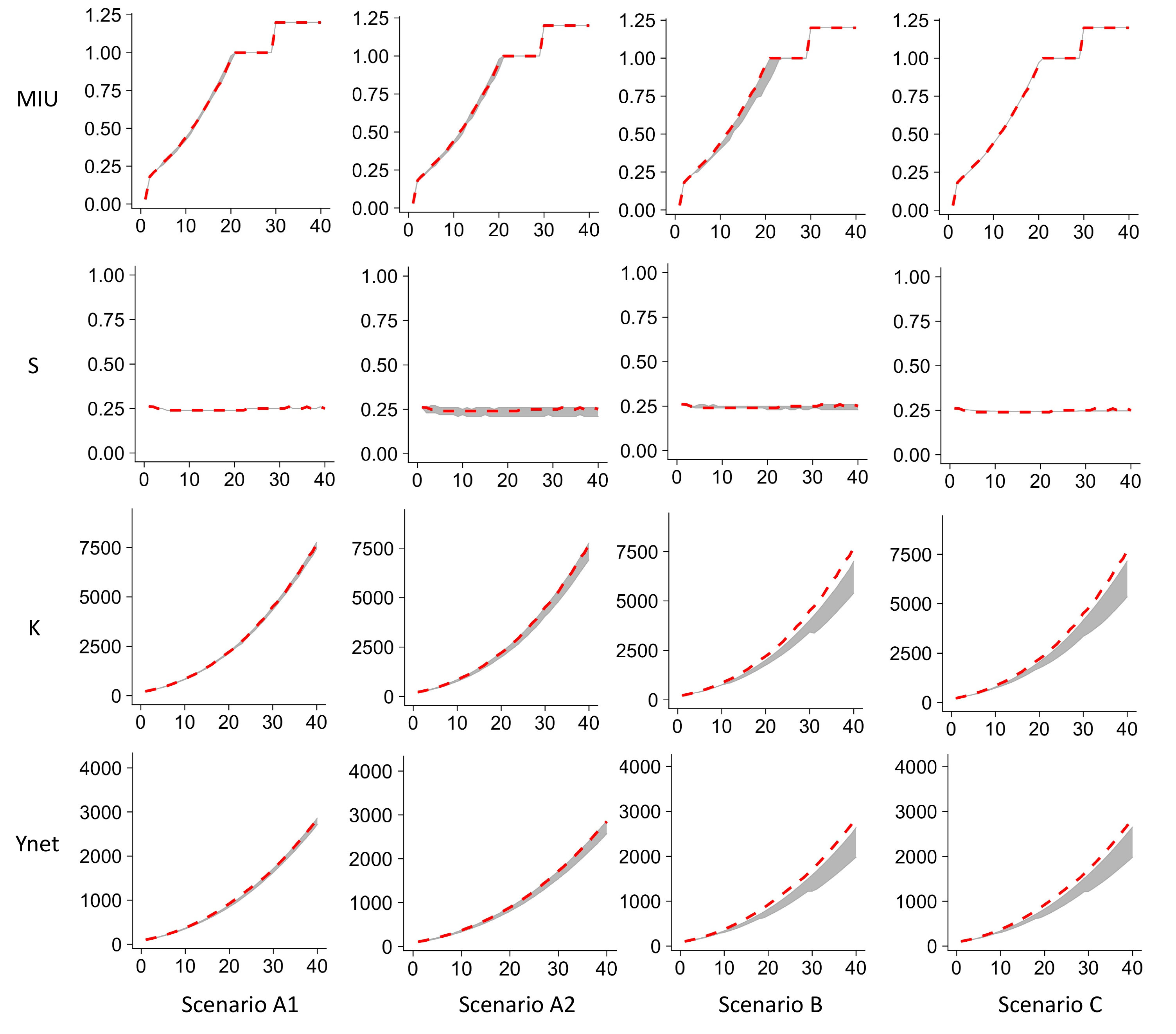}
\captionsetup{width=1.\textwidth}\caption{\footnotesize{Trajectories of optimal solution of DICE model for {\tt MU}, {\tt S}, {\tt K}, {\tt YNET} under various scenarios. The horizontal axes represent time (measured in 5-year steps) from 2015 ($t=0$) to 200 years later ($t=40$). The red dashed lines correspond to the solution of the standard deterministic DICE model (i.e., in the case of no COVID-19 type events), and the gray area represents the 95\% probability interval calculated over 1,000 random trajectories simulated}.}\label{fig1}
\end{figure}

\begin{figure}[p]
\hspace{-1.cm}\includegraphics[scale=0.6]{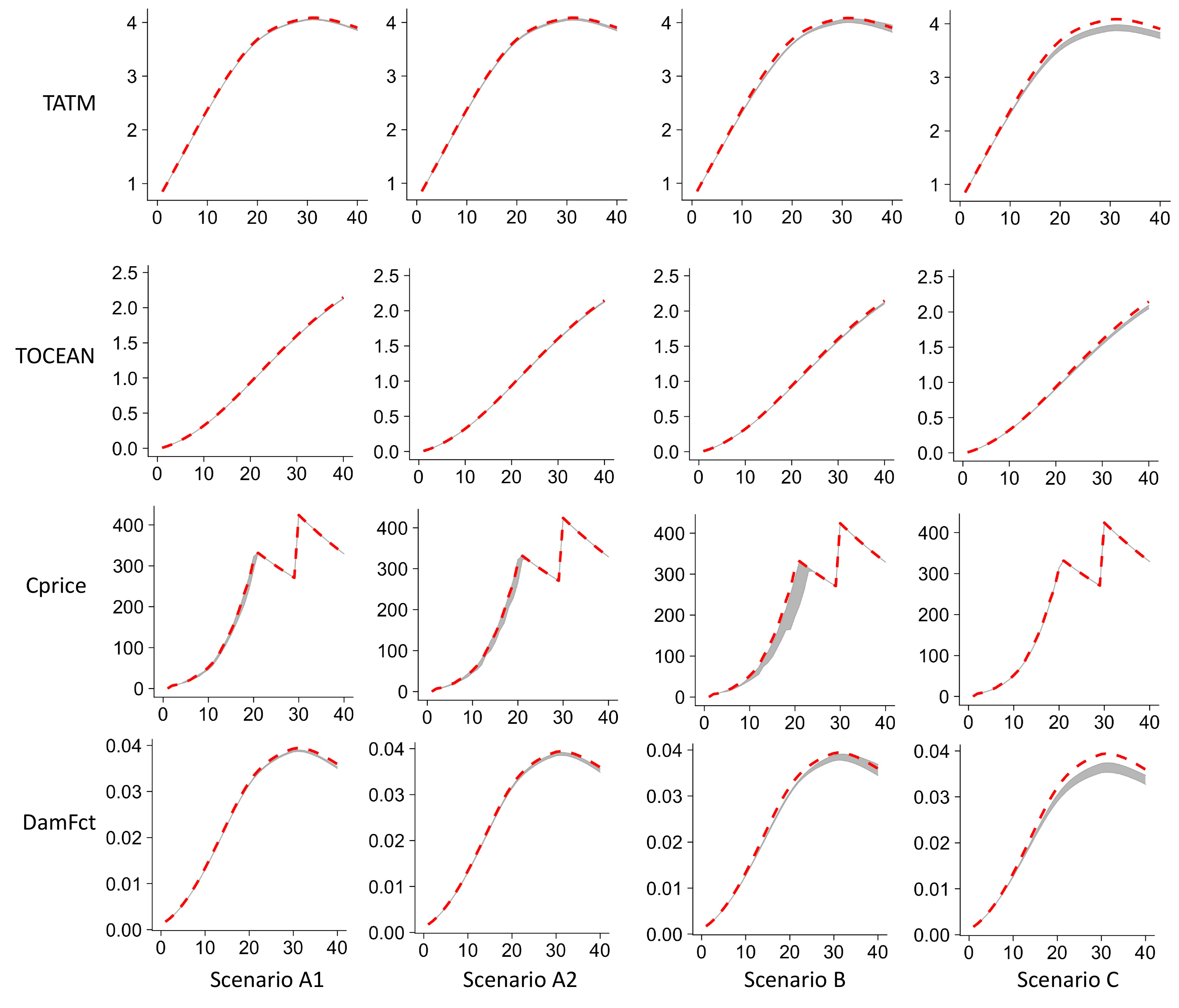}
\captionsetup{width=1.\textwidth}\caption{\footnotesize{Trajectories of {\tt TATM}, {\tt TOCEAN}, {\tt Cprice}, {\tt DamFct}. See the caption of Figure \ref{fig2} for further details.}}\label{fig2}
\end{figure}

\begin{figure}[p]
\hspace{-1.cm}\includegraphics[scale=0.6]{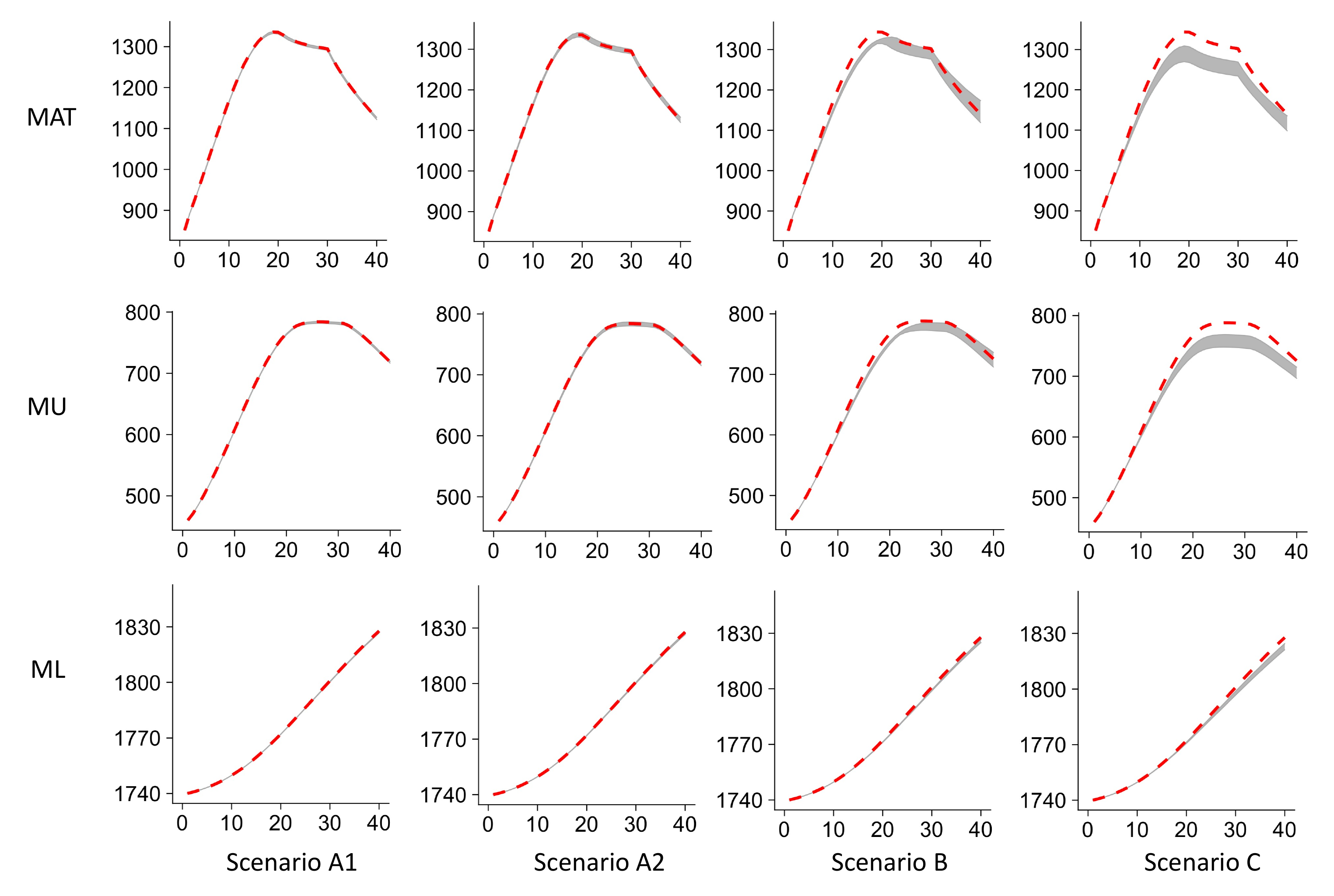}
\captionsetup{width=1.\textwidth}\caption{\footnotesize{Trajectories of {\tt MAT}, {\tt MU}, {\tt ML}. See the caption of Figure \ref{fig1} for further details.}}\label{fig3}
\end{figure}

\noindent Figures \ref{fig1}, \ref{fig2}, and \ref{fig3} show the DICE outputs under the four scenarios described above.
\begin{itemize}
\item Figure \ref{fig1} presents numerical results  for the carbon emission mitigation rate $\mu_t$, savings rate $(1-c_t/Q_t)$, economic capital $K_t$, and the net gross world product $Q_t$, corresponding to plot titles {\tt MIU, S, K, Ynet}.
\item Figure \ref{fig2} plots the results for temperatures $T_t^{AT}$ and  $T_t^{LO}$, carbon price $P_t$, and fraction of output lost owing to a temperature increase $\pi_2(T^{AT}_t)^2$, see equation (\ref{damage_abatement_factor_eq}), corresponding to plot titles {\tt TATM, TOCEAN, Cprice, DamFct}.
\item Figure \ref{fig3} plots the results for carbon concentrations $M^{AT}_t$, $M^{LO}_t$, and $M^{LO}_t$, corresponding to the plot titles {\tt MAT, MU, and ML}.
\end{itemize}


All plots show results under the standard DICE model (i.e., in the case of no random shocks) using the dashed line. For the case of stochastic DICE model, to see the uncertainty/range of outcomes introduced by the shock process $(I_t)_{t\ge 0}$, all plots show the 95\% probability intervals (indicated by the gray area in the plots); these intervals are calculated by simulating 1000 random trajectories and calculating 2.5\% and 97.5\% quantiles over the trajectories at each $t=1,\ldots,N$ to form the interval. These trajectories were simulated forward in time using optimal controls $\mu_t^\ast(\bm{X_t})$ and $c_t^\ast(\bm{X_t})$ obtained by solving the stochastic DICE model, except for scenario C, where controls are taken from the deterministic DICE solution.

The results for Scenario A1 show no material change for all DICE outputs from the deterministic case; only the net output {\tt Ynet} has a small visible gray area owing to stochastic shocks. The gray area is below the deterministic DICE solution for $\tt Ynet$ as expected because shocks reduce net output $Q_t$. The capital state variable $K_t$ also has very small gray area below the deterministic DICE solution consistent with the saving rate {\tt S} virtually unaffected by stochastic shocks while output {\tt Ynet} is reduced by the shocks.

Scenario A2 leads to a more visible (compared to Scenario A1) impact on the economic variables {\tt K} and {\tt Ynet}. The gray area for these variables is below the deterministic solution and corresponds to  approximately 10\% variation in 200 years.  This is expected because the shock size in Scenario A2  is 10\%, larger than 5\% shock under Scenario A1. Saving rate {\tt S} is also slightly but visibly affected with most of trajectories below the deterministic case. 
However, the impact on climate variables (temperature, carbon concentration, emission control rate) from shocks in this scenario  is immaterial. Though, we still can note a very small decrease is emission control rate {\tt MIU} and as a result a small decrease in carbon price {\tt Cprice}; also a tiny drop in {\tt TATM} and in resulting {\tt DamFct}. This is also not surprising because shocks reduce not only the net output but the carbon emission too.

Scenario B clearly leads to larger and material impacts on economic variables {\tt K} and {\tt Ynet} compared to Scenarios A1 and A2 because shocks on the economic output are persistent (propagate recursively to all subsequent time periods). Under Scenario B, there is a material impact on the emissions control {\tt MIU} and carbon price {\tt Cprice}, a small but visible impact on the temperature {\tt TATM}, and a small but visible impact on the concentrations {\tt MAT} and {\tt MU}. Other variables, such as {\tt ML} and {\tt TOCEAN}, are not affected. There is a small impact on the savings rate {\tt S}, which is larger than under Scenario A1, but smaller than that under Scenario A2. The gray area of stochastic DICE trajectories is below deterministic DICE solution for most of the trajectories across all plotted variables. More specifically, stochastic trajectories of  {\tt MIU} and {\tt Cprice} are always below those under the deterministic DICE solution between now and to about 100 years (corresponding to approximately 25\% range for drop of {\tt Cprice} in about 100 years); then, for longer time horizons, there is no difference between trajectories of these variables and their solutions from the deterministic DICE. This means that if we account for stochastic persistent shocks, then the policy for a carbon emission reduction can be less demanding compared to the deterministic case. This is because persistent shocks reduce emission (leading to a reduction in concentration and temperature)  more than simple shocks in Scenarios A1 and A2. Trajectories for {\tt K} and {\tt Ynet} are also below those under the deterministic DICE which is explained by persistent shocks on the net output. In the case of atmospheric temperature {\tt TATM}, and concentrations {\tt MAT} and {\tt MU}, most of the gray area is below the deterministic DICE solution, though there are few trajectories going a bit above deterministic solution after approximately 200 years only.

Finally, the results for Scenario C show the case of applying decisions {\tt MIU} and {\tt S} from the deterministic DICE model to the trajectories of stochastic DICE model. That is, we apply suboptimal decisions that are optimal under the deterministic DICE but suboptimal under the stochastic DICE. Thus, one can see that all trajectories under the stochastic DICE for {\tt MIU}, {\tt Cprice} and {\tt S} are the same as the deterministic DICE solution. Trajectories for all other variables are always appear to be below corresponding deterministic DICE solution. Under this scenario, we see even more material impact on temperature; all trajectories for {\tt TATM} in the gray area are below the deterministic DICE solution. On average, {\tt TATM}  is approximately $0.25^\circ \mathrm{C}$ below the deterministic DICE solution when temperature peaks after about 150 years, with gray area corresponding to about $0.15^\circ \mathrm{C}$ range. The same is for carbon concentration {\tt MAT}, where all trajectories are below the deterministic solution (on average a 10\% drop for a concentration at its peak in about 100 years).
In other words, this scenario shows that more ambitious mitigation targets are now feasible at lower costs, or mitigation targets will be achieved faster if the policy is unchanged (i.e. not adapted for environment with stochastic shocks). This is qualitatively the same as the results of the analysis conducted in \cite{meles2020covid}. Again, this outcome is somewhat expected due to persistent shocks reducing not only the net output {\tt Ynet} but also carbon emission.


\section{Conclusion}
\label{sec:conclusion}
In this paper, we studied the impact of the COVID-19 type events on the carbon concentration, temperature, economic capital and other outputs of the DICE model extended to include corresponding stochastic shocks on the world gross annual output.
We solved the extended model under different scenarios as an optimal stochastic control problem, assuming that shock events occur randomly on average once during a 100 period. The results show that if the world gross output recovers in full after each event
, then the impact of the COVID-19 events on the temperature and carbon concentration will be immaterial even in the case of a conservative 10\% decrease in the annual gross output over a 5-year period.
The impact becomes noticeable but small (the long-term temperature in the atmosphere drops on average by $0.1^\circ \mathrm{C}$) if a 5\% decrease in the gross output owing to a shock is allowed to propagate over time (i.e., allowed to be a persistent shock).
Finally, if the deterministic DICE model policy is still applied in the case of stochastic shocks (i.e., it is a suboptimal policy in this case), then the drop in temperature will be larger (approximately $0.25^\circ \mathrm{C}$). That is, the lower economic activities owing to the occurrence of a shock imply that more ambitious mitigation targets are now feasible at lower costs, which is qualitatively consistent with the results presented in \cite{meles2020covid}.

 Shocks considered in our study reduce both the world net output and emission through the shock reduction of the gross output while tipping point modeling studies such as \cite{lontzek2015stochastic, cai2012dsice} assume shock on the net output only and no shock on the emission.  Thus our shocks lead to a reduction in a policy stringency while tipping point shocks lead to the opposite effect.


In general, incorporation of uncertainty in integrated climate-economy assessment models, such as the DICE model, is an under-developed research topic. Typically, the uncertainty is assessed by calculating the models under the perturbed parameters, and state-of-art stochastic control methods are not really used. This can be partly explained by the difficulty of implementing dynamic programming algorithms. This is probably due to the large number of state variables that call for the use of Monte Carlo simulation methods while until recently, Monte Carlo techniques have not been used for optimal control problems that involve controlled processes. A relatively recent development in this area is the least-squares Monte Carlo approach with the control randomization technique developed in \cite{Kharroubi2014}. However, to solve stochastic control problems maximizing the expected utility (as in the DICE model), some special adjustments are required for this technique, as discussed in \cite{Andreasson2017LSMC}. Implementing this approach for the DICE model incorporating stochasticity in all state variables is the subject of our research project in progress.

Of course, although the DICE model is a typical reference point for many climate-economy studies, it is important to remember that IAMs and the DICE models in particular have significant limitations (in the model structure and model parameters) and have been criticized and debated in the literature (for example, see the discussions in \cite{ackerman2009limitations,pindyck2020use,grubb2021modeling,weitzman2011fat}.

\section{Acknowledgement}
We wish to thank Yoshiki Yamagata for the valuable discussions and comments. Pavel Shevchenko and Tor Myrvoll acknowledge travel support from the Institute of Statistical Mathematics, Japan. We also thank the participants of the 11th conference on the \emph{Transdisciplinary Federation of Science and Technology} for their helpful comments on our preliminary results in \cite{tomokopaveldaisuketor2020}.

\newpage
\bibliographystyle{te}
\bibliography{bibliography}

\end{document}